\documentclass[USenglish]{article}	

\usepackage[utf8]{inputenc}				
\usepackage[big,online]{dgruyter}	
\usepackage{lmodern} 
\usepackage{microtype}
\usepackage[numbers,square,sort&compress]{natbib}
\usepackage{graphicx}
\usepackage{hyperref,graphicx}

\theoremstyle{dgthm}

\theoremstyle{dgdef}

\begin{document}

	\articletype{Research Article}
	\received{Month	DD, YYYY}
	\revised{Month	DD, YYYY}
  \accepted{Month	DD, YYYY}
  \journalname{De~Gruyter~Journal}
  \journalyear{YYYY}
  \journalvolume{XX}
  \journalissue{X}
  \startpage{1}
  \aop
  \DOI{10.1515/sample-YYYY-XXXX}

\title{Dispersive bands of bound states in the continuum}
\runningtitle{Dispersive bands of bound states in the continuum}

\author*[1]{Stefano Longhi}
\runningauthor{S.~Longhi}
\affil[1]{\protect\raggedright 
 Dipartimento di Fisica, Politecnico di Milano, Piazza L. da Vinci 32, I-20133 Milano, Italy  \& IFISC (UIB-CSIC), Instituto de Fisica Interdisciplinar y Sistemas Complejos - Palma de Mallorca, Spain
 
 e-mail: stefano.longhi@polimi.it}
	
	
\abstract{Bound states in the continuum (BICs), i.e. highly-localized modes with energy embedded in the continuum of radiating waves, have provided in the past decade a new paradigm in optics and photonics, especially at the nanoscale, with a range of applications from nano photonics to optical sensing and laser design. Here we introduce the idea of a crystal made of BICs, in which an array of BICs  are indirectly coupled via a common continuum of states resulting in a tight-binding dispersive  energy miniband embedded in the spectrum of radiating waves. The results are illustrated for a chain of optical cavities side-coupled to a coupled-resonator optical waveguide (CROW) with non-local contact points.}


\maketitle

\section{Introduction} 
Bound states in the continuum (BICs), originally predicted in non-relativistic quantum mechanics for certain exotic potentials sustaining localized states with energies embedded in the continuous spectrum of scattered states \cite{r1,r2,r3}, have attracted an increasing interest in optics and photonics over the past decade \cite{r8,r9,r10,r11,r12,r13,r14,r15,r16,r17,r19,r20,r21,r22,r23,r24,r25,r26,r27,r28,r29,r30,r31,r31b,r32,r33,r34,r35,r36,r37,r37a,r38,r39,r40,r41,r41b,r42,r43}, providing a new paradigm for unprecedented light localization in nanophotonic structures (for recent reviews see  \cite{r4,r5,r6,r7,r7b}).  Besides of fundamental interest, BICs have found many interesting applications in several areas of photonics, including integrated and nanophotonic circuits\cite{r23,r39,r43,r6}, laser design \cite{r31,r31b,r32,r41b}, optical sensing \cite{r27,r37a} and nonlinear optics \cite{r21,r34,r40}. Bound states in the continuum have found increasing interest also in cavity and circuit quantum electrodynamics, where spontaneous emission and decoherence can be prevented by the formation of photon-atom BICs \cite{r44a,r44,r45,r46,r47,r48,r49,r50,r51}. 
 Among the different mechanisms underlying the formation of BICs \cite{r1}, we mention symmetry-protected BICs, BICs via separability, Fano or Fabry-P\`erot BICs,  and BICs from inverse engineering. In the majority of cases, a BIC arises from perfect destructive interference of distinct decay channels in the continuum of radiation modes, and thus perturbations rather generally transform a BIC into a long-lived resonance state of the system, so-called quasi-BICs. While the main interest on BICs and quasi-BICs has been focused to their localization features, such as the exceptionally high $Q$ factors achievable using quasi BICs, and to the narrow Fano-like resonances arising from engineering BICs  with applications to optical sensing, a less explored question is the coupling of BICs and related transport features. Arrays of BICs with compact support provide an example of a flat band system, and have been investigated in some previous works \cite{r52,r52b,r53}. However, in such a geometrical setting the BICs are decoupled and transport in the system via BIC hopping is prevented. Coupling of two BICs via a common continuum results rather generally into the formation of two quasi-BICs, which can sustain long-lived Rabi oscillations \cite{r50,r54}. Such results stimulate the search for dispersive bands of BICs, beyond the flat band regime, where transport is impossible.\\
 In this article we introduce the idea of a crystal of BICs where a dispersive band, formed by indirect coupling of a chain of BIC states, is embedded in the broader continuum of radiating waves. Contrary to the BIC flat band systems \cite{r52,r52b,r53}, in our setting the band formed by the BIC states is dispersive, thus allowing transport in the system when e.g. a gradient field is applied. The concept of BIC crystal and transport via BIC mode hopping is exemplified by considering an array of optical cavities side-coupled to a coupled-resonator optical waveguide (CROW) \cite{r55,r56} with non-local contact points.

\section{Dispersive bands formed by indirectly-coupled BICs: Effective non-Hermitian description}
To highlight the idea of a crystal of BICs, let us consider a rather general model describing $N$ discrete states coupled to a common one-dimensional continuum of radiating waves into which they can decay. In photonics, this system can describe, for example, an array of $N$ optical cavities equally spaced by a distance $d$, with the same resonance frequency $\omega_0$, side-coupled by evanescent field to a waveguide, as schematically shown in Fig.1A. A possible photonic platform, suggested in a pioneering work on photonic BICs, could be a 1D photonic crystal waveguide on a square lattice of dielectric rods with additional lateral defects, which displays BICs with low radiation losses \cite{r9}.
 Our main focus is to consider a crystal of side-coupled resonators in the $N \rightarrow \infty$ limit. We note that in cavity or circuit quantum electrodynamics (QED) context the system of Fig.1A can describe as well an array of two-level quantum emitters with transition frequency $\omega_0$, equally-spaced by a distance $d$ and non-locally coupled  by electric-dipole transition to the photon field of waveguide modes. Such circuit QED model could  be implemented, for example, using arrays of artificial atoms side-coupled to a quantum waveguide realized with superconducting quantum circuits \cite{r48}.
 In the following, we will explicitly refer to the photonic system  model and will use the terminology of integrated photonics.\\
In the second-quantization framework, the full Hamiltonian of the photon field is described by Friedrichs-Lee (or Fano-Anderson) Hamiltonian (see for instance \cite{r56c,r56b})
\begin{equation}
\hat{H}=\hat{H}_s+\hat{H}_c+\hat{H}_i
\end{equation}
where 
\begin{equation}
\hat{H}_s= \sum_n \omega_0 \hat{a}^{\dag}_n \hat{a}_n
\end{equation}
is the Hamiltonian of photon field in the $N$ uncoupled cavities with the same resonance frequency $\omega_0$ ($n=1,2,...,N$), 
\begin{equation}
\hat{H}_c= \int dk\;  \omega(k) \hat{c}^{\dag}(k) \hat{c}(k)
\end{equation}
in the Hamiltonian describing the radiating field in the one-dimensional waveguide, with dispersion relation $\omega(k)$ parametrized by the wave number $k$, and
\begin{equation}
\hat{H}_i= \sum_n \int dk \left\{ G_n(k) \hat{a}^{\dag}_n \hat{c}(k)+ {\rm H.c.} \right\}
\end{equation}
describes the resonator-waveguide coupling with spectral coupling functions $G_n(k)$. In the above equations, $\hat{a}_n^{\dag}$ and $\hat{c}^{\dag}(k)$ are the bosonic creation operators of photons in the resonator mode of the $n$-th cavity and in the radiating field of wave number $k$ in the waveguide, respectively. Such operators  satisfy the usual bosonic commutation relations $ [\hat{a}_n, \hat{a}_l^{\dag}]=\delta_{n,l}$, $ [\hat{a}_l, \hat{c}^{\dag}(k)]=0$, $[ \hat{c}(k), \hat{c}^{\dag}(k^{\prime})]= \delta(k-k^{\prime})$, etc. 
 \begin{figure}
\includegraphics[width=15.5cm]{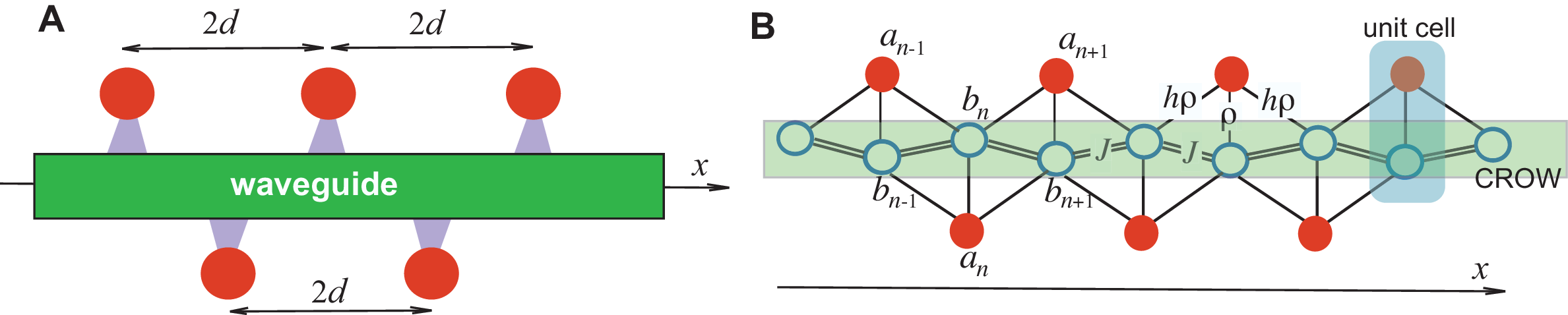}
\caption{{\bf A} Schematic of a waveguide with $N$ side-coupled optical resonators or cavities ($N=5$ in the figure). Adjacent cavities are spaced by a distance $d$. {\bf B} A CROW in a zig-zag geometry (blue rings) with side-coupled optical cavities (red circles). The solid bonds show the couplings of the various cavities/resonators. Note that each cavity is coupled to three resonators of the CROW with coupling constants $ \rho$ and $h \rho$. The distance between adjacent cavities is $d=1$ (in units of the CROW period). The overall cavities-CROW system can be viewed as a bipartite lattice with two sites in the unit cell (sublattice A formed by the side cavities and the sublattice B formed by the resonators of the CROW). The weak-coupling limit corresponds to $\rho \ll J$, where $J$ is the coupling rate of adjacent resonators in the CROW:}
\end{figure}
Assuming time-reversal symmetry, the dispersion relation $\omega(k)$ of the propagating modes in the waveguide satisfies the condition $\omega(-k)=\omega(k)$. We also assume that $\omega_0$ is embedded in the continuous spectrum of waveguide modes and that the resonance condition 
\begin{equation}
\omega(k)=\omega_0
\end{equation}
is satisfied for $k= \pm k_0$, with $\omega(k) \simeq |k| v_g$ for $ k \sim \pm k_0$, where $v_g$ s the group velocity of the left- and right-propagating modes of the waveguide at the frequency $\omega_0$. Since the resonators are equally-spaced by a distance $d$ along the waveguide axis $x$ and equally coupled to the waveguide, the spectral coupling functions $G_n(k)$ satisfy the condition \cite{r10}
\begin{equation}
G_n(k)=G_0(k) \exp(ikdn).
\end{equation}
with $G_0(k) \rightarrow 0$ as $k \rightarrow \pm \infty$.
Let us assume that the system is initially excited by a single photon or by a classical state of light in the resonators solely. In this case we can restrict the analysis to the single excitation sector of Fock space \cite{r49,r57} by letting 
\begin{equation}
 | \psi(z) \rangle  =  \left\{ \sum_n a_n(t) \hat{a}_n^{\dag} | 0 \rangle + \int  dk \; c(k,t) \hat{c}^{\dag}(k) |0 \rangle \right\}  \exp(-i \omega_0 t)
\end{equation}
for the state vector of the photon field, 
where the amplitude probabilities $a_n(t)$ and $c(k,t)$ satisfy the classical ($c$-number) coupled-mode equations
\begin{eqnarray}
i \frac{ da_n}{dt} & = & \int dk \; G_{n}(k) c(k,t)  \\
i \frac{\partial c}{\partial t} & = &  \left\{ \omega(k)-\omega_0 \right\} c(k,t)+ \sum_n G_n^*(k)a_n (t) 
\end{eqnarray}
with $c(k,0)=0$.  Assuming a weak resonator-waveguide coupling and neglecting retardation effects, using standard methods one can eliminate the waveguide degrees of freedom $c(k,t)$ from the dynamics, resulting in an effective non-Hermitian coupling among the resonators mediated by the continuum of waveguide modes  \cite{r9,r10,r7b,r56c,r56b,r58,r58b,r59}.
The resulting approximate non-Hermitian dynamics of the resonator field amplitudes $a_n$ read
\begin{equation}
i \frac{da_n}{dt} \simeq \sum_l H_{n,l} a_l
\end{equation}
 where the elements of the non-Hermitian matrix $H$ are given by (technical details are presented in Sec.1 of the Supplementary Material)
 \begin{equation}
 H_{n,l}=\lim_{\epsilon \rightarrow o^+} \int dk \frac{G_n(k)G_l^*(k)}{i \epsilon+\omega_0-\omega(k)}.
 \end{equation}
 Note that, from Eqs.(6) and (11) it follows that $H_{n,l}$ is a function of $(n-l)$ solely, i.e. 
 \begin{equation}
  H_{n,l}=H_{n-l}= \lim_{\epsilon \rightarrow o^+} \int dk \frac{|G_0(k)|^2  \exp[ikd(n-l)] }{i \epsilon+\omega_0-\omega(k)}.
  \end{equation}
 A few general properties of the coefficients $H_n$ are discussed in Sec.1 of the Supplementary Material. In particular, for a symmetric coupling of the resonator mode with  forward and backward propagating modes of the waveguide, i.e. for $G_0(-k)=G_0(k)$, one has $H_{-n}=H_{n}$. Moreover, $H_n$ vanishes as $n \rightarrow \infty$ if and only if the condition
 \begin{equation}
 G( \pm k_0)=0
 \end{equation}
 is satisfied. As shown in the Supplementary Material, this condition corresponds to vanishing of the imaginary part of $H_0$, i.e. to the existence of a BIC when \emph{a single} resonator (rather than a chain of resonators) is coupled to the waveguide.\\ 
   In a system with discrete translation invariance, i.e. in the $N \rightarrow \infty$ limit, Eq.(10) indicates that the indirectly-coupled resonators behave like a tight-binding crystal with long range hopping, sustaining Bloch-like supermodes of the form
   \begin{equation}
   a_n= A \exp[iqdn-i \Omega(q)t]
   \end{equation}
    with the energy dispersion relation given by
     \begin{equation}
 \Omega(q)=\sum_n H_{n} \exp(-iqdn)
 \end{equation}
 where $-\pi/d \leq q < \pi/d$ is the Bloch wave number.
 The main result is that, provided that the condition (13) is met, the series on the right hand side of Eq.(15) converges and the energy spectrum, described by the dispersion curve $\Omega(q)$, is entirely real and given by (see Sec.2 of the Supplementary Material for technical details)
\begin{equation}
\Omega(q)= \frac{2 \pi}{d}  \sum_{l=-\infty}^{\infty} \frac{|G_0(q+ 2 \pi l/d)|^2}{\omega_0-\omega(q+ 2 \pi l /d)}.
\end{equation}
 Clearly, the dispersion curve $\Omega(q)$ s embedded in the broad spectrum $\omega(k)$ of scattering states of waveguides. A trivial case is when the BIC modes are compact states  and decoupled one from another ($H_n=0$ for $n \neq 0$), which occurs for enough wide spacing $d$: in this case one obtains a {\em flatband} BIC crystal, i.e.  $\Omega(q)=H_0$ independent of $q$. Examples of crystals sustaining a flat band of BIC states have been discussed in the context of flat band systems (see e.g. \cite{r52,r52b,r53}).  However, the present asymptotic analyst shows that an entirely real energy dispersion relation is possible beyond the flat band case when the BIC states are indirectly coupled via the continuum of the waveguide modes, so that excitation can be transferred among the various indirectly-coupled BICs through the embedded dispersive band $\Omega(q)$. 
 
 \section{An example of a BIC lattice: optical cavities side-coupled to a coupled-resonator optical waveguide (CROW)}
 \subsection{Model and band structure}
To illustrate the existence and properties of a BIC crystal with a dispersive band, we consider a simple and exactly-solvable tight-binding model, shown in Fig1B. It consists of  an array of optical cavities side-coupled with three contact points to a CROW in a zig-zag geometry, with spacing $d=1$ between one cavity and the next one  (in units of the CROW period). We assume negligible radiation losses from the cavities to the surrounding space, so that  decay mainly arise from evanescent mode coupling to the CROW. This system could be physically implemented in a 2D photonic crystal platform, where a periodic sequence of defect modes is side-coupled to a 1D photonic crystal waveguide, in a configuration similar to the case of the two off-channel defects suggested in Ref.\cite{r9}. In such a configuration the  side-coupled off-channel cavities are hide deeply in the photonic crystal so that radiation losses from the cavities are strongly suppressed. For the case of $N=2$ cavities, this model has been recently studied in Ref.\cite{r54}, showing that weakly damped Rabi flopping can be observed owing to the indirect coupling of two quasi-BIC modes. Here we consider, conversely, the $N \rightarrow \infty$ limit so that the system displays discrete translation invariance and we can apply Bloch  band theory. We indicate by $\omega_0$ the frequency detuning between the optical modes in the cavities and resonators, by $J$ the coupling constant between adjacent resonators in the CROW, and by $\rho$ and $h \rho$ the coupling of each optical cavity with the three closest resonators of the CROW (Fig.1B). The dimensionless parameter $h$ measures the relative strength of nearest and next-to-the-nearest mode coupling of the optical cavity with the CROW resonators.  We typically assume $\rho<J$, with $\rho \ll J$ in the weak coupling regime. Clearly, in the $N \rightarrow \infty$ limit the tight-binding Hamiltonian in Wannier basis can be readily solved by Bloch theorem and the energy spectrum consists of two bands since the entire system can be viewed as a bipartite lattice composed by two sublattices (Fig.1B). This analysis will be discussed below. Before, we wish to illustrate how the existence of the BIC crystal can be predicted from the effective non-Hermitian description outlined in the previous section. As shown in Sec.3 of the Supplementary Material, the Hamiltonian of the photon field for the model of Fig.1B can be cast in the general Friedrichs-Lee form of Eqs.(1-4), where the continuum of radiation modes in the waveguide are the Bloch modes of the CROW with the dispersion relation 
 \begin{figure}
\includegraphics[width=15.5cm]{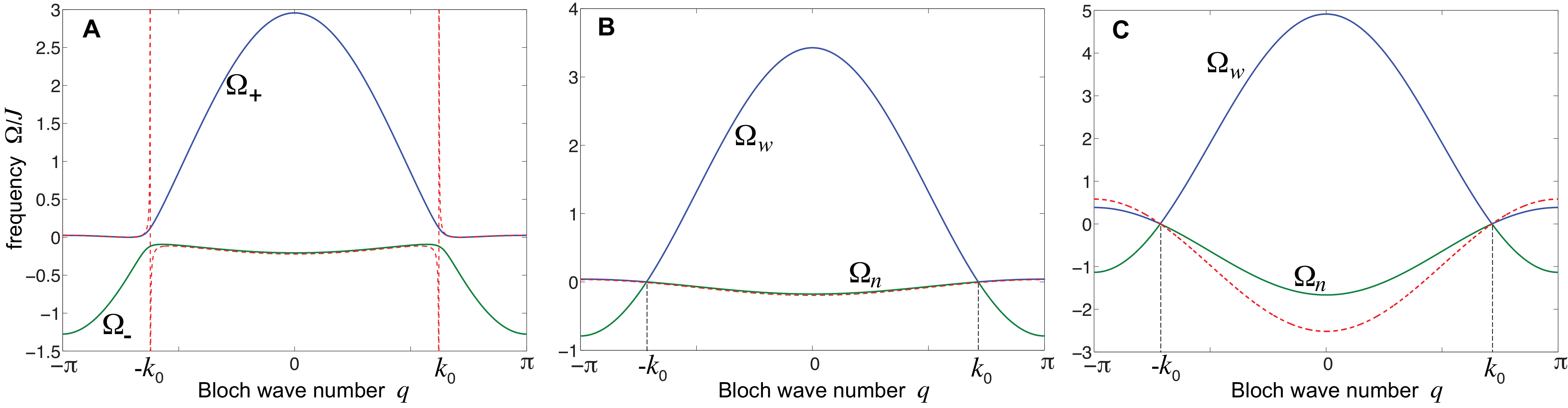}
\caption{{\bf A,B.} Band diagram of the CROW structure with side-coupled optical cavities of Fig.1B for parameter values $\rho/J=0.3$, $h=0.8$ and for {\bf A} $\omega_0/ J=-0.75 $, {\bf B} $\omega_0 /J=-1/h=-1.25$. The dashed red curves in {\bf A} and {\bf B} show the band dispersion curve $\Omega(q)$ predicted by the asymptotic analysis [Eq.(20)]. In {\bf A} the BIC condition $\omega_0=-J/h$ is not satisfied, the spectrum is gapped with non-crossing upper ($\Omega_+$) and lower ($\Omega_-$) bands. Note that $\Omega(q)$ shows a singularity at the Bloch wave numbers $ q= \pm k_0$, where $k_0$ satisfies the resonance condition $\omega_0=2 J \cos k_0$. In {\bf B} the BIC condition $\omega_0=-J/h$ is satisfied and the spectrum is gapless, resulting in the formation of narrow band $\Omega_n$ (the BIC band) embedded in a wider band $\Omega_w$. In this case the predicted curve $\Omega(q)$ is almost overlapped with the exact curve $\Omega_n$ of the narrow band. {\bf C.} Band diagram in the strong coupling regime ($\rho/J=1.1$, $h=0.8$) with $\omega_0/J=-1/h$. A narrower band $\Omega_n$ embedded in a wider band $\Omega_w$ is still observed in the exact model (solid curves). However, the asymptotic analysis [Eq.(21), dashed curve] fails to provide the accurate shape of the narrow band.}
\end{figure}
\begin{equation}
\omega(k)=2 J \cos k,
\end{equation}
[$k$ varies in the range $(-\pi, \pi)$], while the spectral coupling function $G_0(k)$ is given by
\begin{equation}
G_0(k)= \frac{\rho}{ \sqrt{2 \pi}} \left( 1+2 h \cos k  \right). 
\end{equation}
The BIC condition (13) is satisfied provided that the frequency detuning $\omega_0$ is tuned to the value
\begin{equation}
\omega_0= 2 J \cos k_0=-J/h
\end{equation}
so as $G( \pm k_0)=0$. Note that, to satisfy Eq.(19), we necessarily require $h>1/2$.\\ 
The dispersion curve of the BIC crystal is obtained from Eq.(16), where the sum on the right hand side of the equation is limited to the $n=0$ term solely (this is because $k$ varies only in the range $(-\pi, \pi)$, rather than from $-\infty$ to $\infty$ as in a waveguide). One obtains 
\begin{equation}
\Omega(q)=\frac{\rho^2 (1+2 h \cos q)^2}{\omega_0-2J \cos q}
\end{equation}
which shows a singularity at $ J \cos q = \omega_0 /2$. The physical origin of such a singularity and failure of the asymptotic method will be discussed below through the exact analysis. However, the curve $\Omega(q)$ is regularized under the BIC condition $\omega_0=-J/h$ [Eq.(19)], yielding
\begin{equation}
\Omega(q)=-\frac{\rho^2 h}{J} (1 +2 h \cos q)
\end{equation}
Note that the dispersion curve is sinusoidal, corresponding to the vanishing of $H_n$ for $n \neq 0, \pm 1$.\\
Let us then turn to the \emph{exact} analysis of modes and spectrum of the lattice shown in Fig.1B, which can be readily obtained by standard Bloch analysis in the Wannier basis of the CROW modes.
In fact, indicating by $a_n$ and $b_n$ the field amplitudes in the $n$-th side-coupled optical cavity and in the $n$-th resonator of the CROW, the classical coupled-mode equations for $a_n$ and $b_n$ read
\begin{eqnarray}
i \frac{da_n}{dt} & = & \omega_0 a_n +\rho b_n + \rho h (b_{n-1}+b_{n+1}) \\
i \frac{db_n}{dt} & = & J(b_{n+1}+ b_{n-1})+ \rho a_n+ \rho h (a_{n+1}+a_{n-1}).
\end{eqnarray}
The Bloch eigenstates (supermodes) of the lattice are of the form
\begin{equation}
\left(
\begin{array}{c}
a_n \\
b_n
\end{array}
\right)= \left(
\begin{array}{c}
A \\
A
\end{array}
\right) \exp[iqn-i \omega_0t -i \Omega(q) t]
\end{equation}
where the dispersion relation $\Omega(q)$ is obtained as the root of the determinantal equation
\begin{equation}
\left|
\begin{array}{cc}
\Omega(q) & -\rho (1+2h \cos q) \\
-\rho (1+2h \cos q) & \Omega(q)+\omega_0-2 J \cos q
\end{array}
\right|=0.
\end{equation}
Note that Eq.(25) is of second order in $\Omega(q)$, indicating the existence of two lattice bands, according to the bipartite nature of the lattice which comprises the two sublattices A (optical cavities) and B (resonators of the CROW). The dispersion curves of the two bands read
\begin{equation}
\Omega_{\pm}(q)=-\frac{\omega_0}{2}+J \cos q \pm \sqrt{ \left( \frac{\omega_0}{2}-J \cos q \right)^2 +\rho^2 (1+2 h \cos q)^2}.
\end{equation}
 Rather generally the energy spectrum is gapped with an upper ($\Omega_+$) and lower ($\Omega_-$) band (Fig.2A). However, when the resonance frequency detuning $\omega_0$ is tuned to the value $\omega_0=-J/h$, corresponding to the BIC condition (19), the spectrum is gapless and the two bands cross, leading to the formation of a narrow band $\Omega_n(q)$, embedded in a wider band $\Omega_w(q)$, as shown in Fig.2B. Basically, $\Omega_n(q)=\Omega_-(q)$ for $-k_0 \leq q \leq k_0$, and $\Omega_n(q)=\Omega_+(q)$ for $|q|>k_0$. Likewise, $\Omega_w(q)=\Omega_+(q)$ for $-k_0 \leq q \leq k_0$, and $\Omega_w(q)=\Omega_-(q)$ for $|q|>k_0$.
Remarkably, the narrow and wider bands have the same dispersion relation but with different amplitudes $\Delta_{w.n}$, namely
\begin{equation}
\Omega_{w.n}(q)= \Delta_{w,n} (1+2 h \cos q)
\end{equation} 
 with 
 \begin{equation}
 \Delta_{w,n}=\frac{J}{2h} \pm \sqrt{\left( \frac{J}{2h} \right)^2+ \rho^2}.
 \end{equation}
Note that, in the weak coupling limit $\rho \ll J$ one has $\Delta_n \simeq -h \rho^2/J$, so that the dispersion relation $\Omega_n(q)$ of the embedded narrow band exactly reduces to Eq.(21), predicted by the asymptotic analysis. The corresponding Bloch eigenstates [Eq.((24)] have the largest excitation in the sublattice $a_n$, i.e. in the optical cavities, with a ratio $B/A = \Delta_n / \rho \sim O(\rho /J)$.\\ 
It is worth briefly discussing the failure of the asymptotic analysis when the BIC condition $\omega_0=-J/h$ is not met and the appearance of a singularity in the energy band $\Omega(q)$, given by Eq.(20), at the Bloch wave number $q= \pm k_0$ with  $J \cos k_0= \omega_0/2$. To this aim, let us consider the exact band structure of the system, given by Eq.(26), in the $\rho / J \rightarrow 0$ limit. Let us write the bands $\Omega_{\pm}$ in the equivalent form
\begin{equation}
\Omega_{\pm}(q)= -\frac{\omega_0}{2}+J \cos q \pm \left| \frac{\omega_0}{2}-J \cos q \right|  \sqrt{1 +\frac{\rho^2 (1+2 h \cos q)^2}{\left( \frac{\omega_0}{2}-J \cos q \right)^2} }.
\end{equation}
In the limit $\rho / J \rightarrow 0$ and for $\omega_0 \neq -J/h$, a narrow gap, corresponding to an avoided crossing of the two branches $\Omega_{\pm}(q)$, occurs at the Bloch wave number $q= \pm k_0$ such that $J \cos k_0= \omega_0/2$. For $q$ far enough from $\pm k_0$ such that $|2 J \cos q-\omega_0| \gg \rho$, we can perform a Taylor expansion of the square root on the right hand side of Eq.(29), yielding for the mixed branch $\Omega_n(q)$ the following expression
\begin{equation}
\Omega_n(q) \simeq \frac{\rho^2 (1+2 h \cos q)^2}{ \omega_0-2J \cos q } 
\end{equation}
which is precisely the dispersion curve Eq.(20) predicted by the asymptotic analysis. Hence, the unphysical singularity found in Eq.(20) at the Bloch wave numbers $q= \pm k_0$ is related to the formation of a physical band gap, which cannot be described by the asymptotic analysis.\\
Finally, let us mention that for the exactly solvable model given by Eqs.(22,23) a narrower dispersive band $\Omega_n$ embedded into a wider band $\Omega_w$ is observed even in the strong coupling regime ($\rho$ comparable or even larger than $J$), provided that the gapless condition $\omega_0=-J/h$ is met. An example of the exact band structure ($\Omega_n$ and $\Omega_w$) in the strong coupling regime is shown in Fig.2C by solid curves. Clearly, in this regime the asymptotic analysis, valid in the weak coupling limit, does not provide anymore an accurate description of the embedded band $\Omega_n$, with the dispersion curve $\Omega(q)$ predicted by Eq.(21) (dashed curve in Fig.2C) substantially deviating from the exact band dispersion curve $\Omega_n$. Also, in the strong coupling regime a single BIC mode does not exist anymore under the condition (19), and an additional shift from the gapless value given by Eq.(19) would be required to have a single BIC: hence in the strong coupling regime the nature of the embedded band $\Omega_n$ cannot be readily linked to indirectly-coupled isolated BIC modes of the structure.

 \begin{figure}
\includegraphics[width=15.5cm]{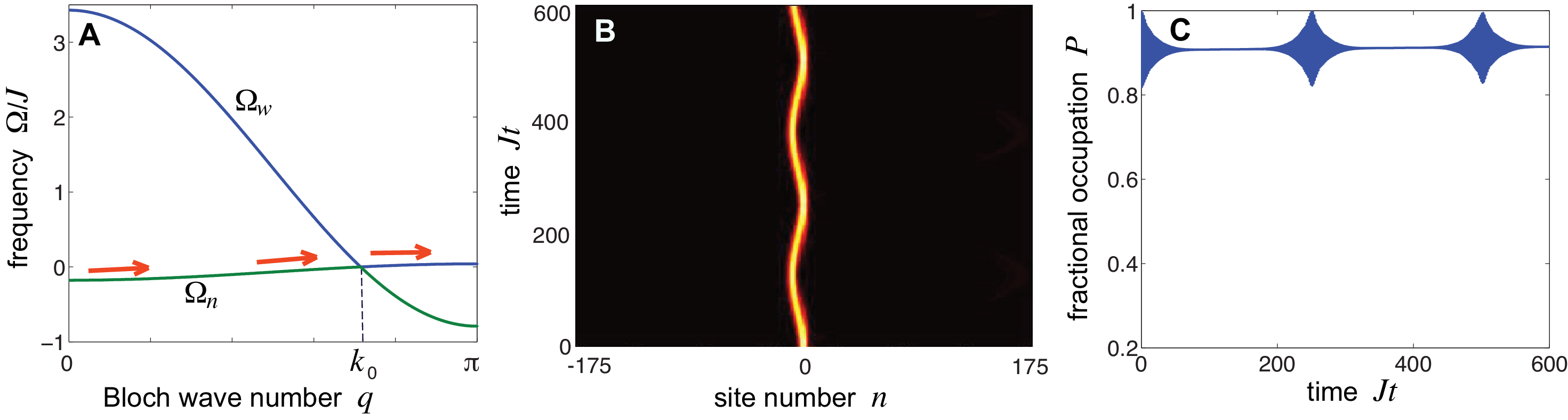}
\caption{Bloch oscillation dynamics in the BIC crystal with the same parameter values as in Fig.2{\bf B} and for a gradient $F/J=0.025$. The initial excitation condition is a Gaussian wave packet $a_n(0)=\exp[-(n/w)^2+iq_0 n]$ with initial wave number $q_0=0$ and width $w=3$. The external gradient $F$ induces a drift of the Bloch wave number according to $q(t)=q_0+Ft$, which is schematically depicted in {\bf A} by solid arrows. When the wave number $q$ reaches the crossing point $q=k_0$, the wide band $\Omega_w$ is not excited and the entire wave packet undergoes a periodic motion, remaining trapped in the narrow (BIC) band $\Omega_n$, with a period $T_B= 2 \pi /F \simeq 251.3 /J$. This is shown in panel {\bf B}, which depicts on a pseudo color map the temporal evolution of the amplitudes $|a_n(t)|$ in the various cavities, labelled by the site number $n$. The temporal evolution of the fractional optical power $P(t)$ trapped in the cavities, defined by Eq.(33), is shown in panel {\bf C}. Note that during the entire BO period the optical power remains mostly trapped in the optical cavities, with small excitation of the resonators in the CROW.}
\end{figure}

\subsection{Transport dynamics: Bloch oscillations in the BIC crystal}
 In a flat band system realized by a chain of uncoupled BICs, transport via mode hopping in the flat band is clearly prevented. Conversely, in a BIC crystal with a dispersive band transport is possible.  To illustrate the concept, let us introduce a linear refractive index gradient along the axis $x$ of the system shown in Fig.1B.In a single-band crystal, such a gradient is known to induce a periodic light dynamics which is the analogous of Bloch oscillations (BOs) of electrons in crystalline potentials (see for instance \cite{B01,B02,B03,B04,B05,B06,B07,B08}). 
 The coupled-mode equations (22,23) are modified as follows
 \begin{eqnarray}
i \frac{da_n}{dt} & = & \omega_0 a_n +\rho b_n + \rho h (b_{n-1}+b_{n+1})+Fna_n \\
i \frac{db_n}{dt} & = & J(b_{n+1}+J b_{n-1})+ \rho a_n+ \rho h (a_{n+1}+a_{n-1})+Fnb_n,
\end{eqnarray}
where $F$ is a small frequency detuning gradient induced by the index gradient. As an initial condition, we assume a Gaussian wave packet excitation of the optical cavities (sublattice A) with vanishing momentum, i.e. $a_n(0)=\exp(-n/^2w^2+i q_0 n)$, $b_n(0)=1$ with $q_0=0$ and $w \gg 1$. The observed dynamical behavior largely depends on whether the system is gapless ($\omega_0=-J/h$) or gapped  ($\omega_0 \neq -J/h$).\\
In the gapless case, i.e. when we have a BIC crystal, in the weak coupling limit $\rho \ll J$ the initial wave packet excites the narrow portion of the $\Omega_n(q)$ narrow band  with wave number $q \sim q_0=0$. In the semiclassical description of the wave packet dynamics, the external force $F$ induces a drift $q(t)=q_0+Ft$ of the wave number (momentum) in Bloch space.  Since the spectrum is gapless, at the crossing points $q= \pm k_0$  the excitation fully remains in the BIC crystal narrow band $\Omega_n$, without excitation of the wider embedding band $\Omega_w$, i.e. of  sites of sublattice B. As a result, a periodic dynamics of the wave packet with a BO period $T_B= 2 \pi /F$ is observed, with most of excitation remaining trapped in the optical cavities, as measured by the evolution of the fractional optical power
\begin{equation}
P(t)= \frac{\sum_n |a_n(t)|^2}{\sum_n |a_n(t)|^2+\sum_n |b_n(t)|^2}.
\end{equation}
This is illustrated, as an example, in Fig.3.\\
\begin{figure}
\includegraphics[width=15.5cm]{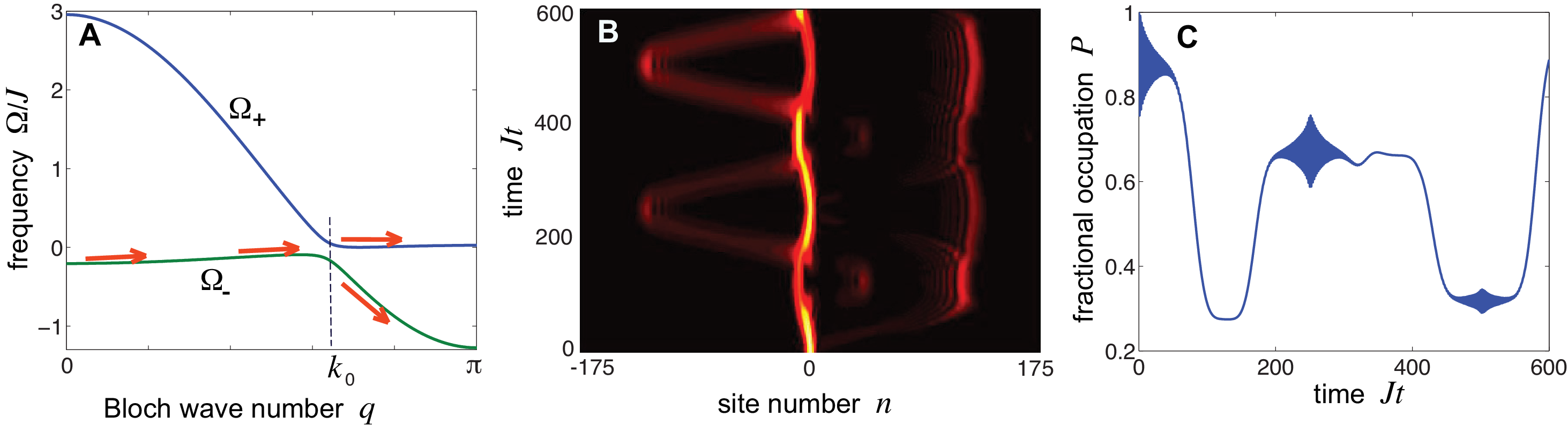}
\caption{Same as Fig.3, but for parameter values as in Fig.2{\bf A} and for a gradient $F/J=0.025$. Note that, when the wave number $q(t)$ of the Gaussian wave packet reaches the crossing point $q=k_0$, excitation is splitted  between the upper and lower bands $\Omega_{\pm}$ (solid arrows in panel {\bf A}), and the entire wave packet undergoes an irregular motion, dubbed Bloch-Zener oscillations (panel {\bf B}), with aperiodic exchange of optical power between the two bands as $q(t)$ crosses the two gap points $q=\pm k_0$. Correspondingly, a large fraction of optical power can be transferred during the dynamics from the optical cavities (sublattice A) to the resonators of the CROW (sublattice B), as shown in panel {\bf C}.}
\end{figure}
Conversely, when the condition $\omega_0=-J/h$ is not satisfied, i.e. when the two bands of the bipartite lattice undergo an avoided crossing at $q= \pm k_0$, we do not not have a BIC crystal anymore, i.e. a narrow band embedded in a wider one is not formed, rather we have two distinct bands $\Omega_{\pm}(q)$ separated by a gap. The initial wave packet mostly excites the Bloch modes of the lower band $\Omega_-(q)$ with wave number near $q \sim q_0=0$. When the drifting Bloch wave number $q(t)=q_0+Ft$ reaches the avoided crossing points $q= \pm k_0$, a splitting of excitation between the two distinct bands $\Omega_{\pm}$ is observed, leading to a more complex and generally aperiodic two-band dynamics (so-called Bloch-Zener oscillations \cite{BZ1,BZ1b,BZ2,BZ3,BZ4}). This behavior is illustrated in Fig.4. Note that, contrary to gapless case of Fig.3, at the avoided-crossing points a rather abrupt change of the fractional optical power $P(t)$ trapped in the optical cavities is observed, i.e. excitation does not remain trapped in the optical cavities anymore, but it is periodically transferred into the resonators of the CROW.

\section{Conclusion}
In this work we have introduced the concept of a BIC crystal, i.e. a narrow energy band of Bloch modes (rather than one or more discrete energy levels) embedded in a wide spectrum of radiation modes. By means of an asymptotic analysis based on a non-Hermitian effective model, it has been shown that rather generally a BIC cryatl arises when single BIC modes are indirectly coupled via a common continuum. To support the predictions of the asymptotic analysis, we presented an exactly-solvable model, consisting of an array of optical cavities side-coupled non-locally to a CROW. The narrow embedded energy band formed by the indriectly-copuled BIC modes enables transport in the crystal, contrary to the case of a flat band formed by uncoupled BIC modes. Our results shed new light in the are of BIC states, suggesting the fresh concept of {\it embedded band} or {\it BIC crystal}, that could stimulate interest in photonics and in other areas of physics, such as in cavity or circuit quantum electrodynamics.


\begin{acknowledgement}
 The author acknowledges the Spanish State Research Agency through the  Severo Ochoa
and Maria de Maeztu Program for Centers and Units of Excellence in R\&D (MDM-2017-0711).
\end{acknowledgement}


\end{document}